# THE SPACE DISTRIBUTION OF THE LYMAN ALPHA CLOUDS IN THE LINE OF SIGHT TO THE Z=3.66 QSO 0055-269 *

*Stefano Cristiani*
Dipartimento di Astronomia, Università di Padova,
vicolo dell' Osservatorio 5, I-35122 Padova, Italy

*Sandro D'Odorico*
European Southern Observatory
Karl Schwarzschild Str. 2, D-85748 Garching, Germany

*Adriano Fontana*
Dipartimento di Fisica, II Università di Roma,
via E. Carnevale, I-00173 Roma, Italy

*Emanuele Giallongo*
Osservatorio Astronomico di Roma,
via dell'Osservatorio, I-00040 Monteporzio, Italy

*Sandra Savaglio*
Dipartimento di Fisica, Università della Calabria
I-87036 Arcavata di Rende, Cosenza, Italy

# ABSTRACT

The spectrum of the quasar Q0055-269 ($z = 3.66$) has been observed at the resolution of 14 km s$^{-1}$ in the wavelength interval 4750–6300 Å. We give a list of Lyman$-\alpha$ lines and metal-line systems for which column densities and Doppler widths have been derived by a fitting procedure. The statistical distribution of the Doppler parameter for the Lyman$-\alpha$ lines is peaked at $b \simeq 23$ km s$^{-1}$, with 13% of lines with $15 \leq b \leq 20$ km s$^{-1}$. The column density distribution of the Lyman$-\alpha$ lines is described by a power-law with a break or cutoff at log $N_{HI} \simeq 14.5$. A featureless power-law distribution is rejected with a probability of 99.94%. Significant clustering, with $\xi \simeq 1$ at $\Delta v = 100$ km s$^{-1}$, is detected only for lines with log $N_{HI} \gtrsim 13.8$. In addition, two voids of size $\sim 2000$ km s$^{-1}$ are found in the spectrum with a random probability of $2 \times 10^{-4}$. From the proximity effect for lines in the interval log $N_{HI} = 13.3 - 14.2$ it is possible to infer an UV ionizing background $J = 5 \times 10^{-22}$ ergs s$^{-1}$ cm$^{-2}$ Hz$^{-1}$ sr$^{-1}$, within a factor of two from the integrated quasar contribution.

*Subject headings*: intergalactic medium – quasars: absorption lines – quasars: individual (0055-269)

## 1. INTRODUCTION

The spectra of quasars shortward of the Lyman$-\alpha$ emission show a dramatic increase of the opacity, mainly due to a forest of narrow Lyman$-\alpha$ absorption lines caused by intervening clouds along the line-of-sight (Lynds 1971; Sargent et al. 1980), with only a small contribution of elements heavier than hydrogen.

Any additional continuous absorption from an intergalactic medium (IGM) (Gunn & Peterson 1965) has proved to be elusive: tight upper limits $\tau < 0.02 \pm 0.03$ have been derived up to the highest observable redshifts, $z_{abs} \lesssim 4.5$ (Giallongo et al. 1992,1994), implying that any smoothly distributed IGM must be highly ionized. A baryon density $\Omega_{IGM} \lesssim 0.01$ is derived assuming a photoionization equilibrium with the UV ionizing background expected from the quasar population. Such a limit, when compared with the total baryon density deduced from the nucleosynthesis ($\Omega_{\rm baryon} \leq 0.05 h_{50}^{-2}$, Walker et al., 1991) implies that a large fraction of the baryons is already in bound systems, e.g. Lyman$-\alpha$ clouds, at $z \sim 5$.

The knowledge of the physical and cosmological properties of the Lyman$-\alpha$ absorbers relies on a few parameters (redshift, column density, Doppler width) which can be derived from high resolution spectroscopy and line profile fitting of often blended absorption features. Typical values of the neutral hydrogen column density and Doppler width seem to depend on the resolution, the deblending procedures and the contamination of very narrow and weak metal lines in the Lyman−$\alpha$ forest. At the resolution of 20-30 km s$^{-1}$ used in the past decade, the derived average value of the Doppler width was $b = 30$ km s$^{-1}$ and the column density distribution was found to be a featureless power-law of the type $f(N_{HI}) \propto N_{HI}^{-1.7}$ for $\log N_{HI} \geq 13.7$ cm$^{-2}$ (Carswell 1988 and references therein). This average Doppler width, corresponding to an average temperature $T \sim 5 \times 10^4$ K, is typical for clouds in photoionization equilibrium with the general UV ionizing background (UVB) produced by quasars and/or galaxies. The inferred models predict tenuous gas envelopes with densities $n_H \sim 10^{-3} - 10^{-4}$ cm$^{-3}$ and sizes $\gtrsim 10$ kpc in a state of high ionization $n_H/n_{HI} \sim 10^4$. The clouds are confined by the pressure of the external diffuse IGM (Sargent et al. 1980, Ikeuchi & Ostriker 1986) or by the gravitational potential of non-baryonic dark halos (Rees 1986, Ikeuchi 1986).

Echelle data at 10-30 km s$^{-1}$ resolution (Pettini et al. 1990, Carswell et al. 1991, Rauch et al. 1992, 1993, Giallongo et al. 1993, GCFT, Fan & Tytler 1994) show also a non negligible number of narrow lines with $b \lesssim 20$ km s$^{-1}$ (i.e. $T \lesssim 24000$ K). Although the reported fraction of narrow lines is wildly variable in these works (from 12% to 63% of the overall sample) and possibly subject to effects due to the low signal-to-noise (see e.g. Rauch et al. 1993), the presence of lines with $b \sim 18 - 20$ km s$^{-1}$ (i.e. $T \gtrsim 19000$ K) could have interesting consequences on photoionization models for the Ly$\alpha$ clouds. If the clouds are photoionized by the UVB produced by quasars, then low temperatures can be obtained increasing the gas density. In this case the cloud sizes are very small ($\sim 1$ pc) and in contrast with the observation of coincident absorptions in gravitationally lensed quasar pairs, that imply lower limits for the diameter of spherical clouds in the range $12-50h_{50}^{-1}$ kpc (Pettini et al. 1990, Smette et al. 1992, Shaver private communication). Adiabatic cooling due to the expansion of the clouds against the external pressure of the diffuse IGM has been invoked to explain such low temperatures together with large sizes (Duncan, Vishniac, & Ostriker 1991; Petitjean et al. 1993a).

Alternatively Giallongo & Petitjean (1994) have proposed a scenario in which the inverse Compton cooling of the cloud electrons on the cosmic microwave background and a strong decrease of the UVB flux just shortward of the HeII edge allow temperatures $\lesssim 20000$ K even for clouds with large sizes $> 20$ kpc and in photoionization equilibrium at $z \geq 3$. The observed temperature distribution in the Lyman−$\alpha$ sample obtained at $z \sim 3$ from observations of PKS 2126-158 (GCFT) is then explained in terms of space fluctuations in the UVB at energies higher than the HeII edge. The clouds result to have low densities $n_H \lesssim 10^{-4}$ cm$^{-3}$ and sizes that for $\log N_{HI} = 14.5$ are of the order of the Jeans length. Lyman−$\alpha$ clouds with $\log N_{HI} > 14.5$ are then gravitationally unstable and disappear in a short time scale. This produces a feature in the column density distribution consistent with the break found in PKS 2126-158 at $\log N_{HI} = 14.8$.

A strong decrease of the UVB flux at the HeII edge is expected on the basis of observations by Jakobsen et al. (1994) of a Gunn-Peterson HeII absorption trough at $z \sim 3$ in the HST spectrum of the quasar Q0302-003.

If gravity becomes more and more important for clouds with $\log N_{HI} \gtrsim 14$, some clustering in the space distribution of these lines is expected. To gain sufficient statistics to study the two point correlation function of those lines, a large sample in each QSO spectrum is required, prompting to observations of quasars at high emission redshifts $z_{em} > 3$.

Thus, to study the physical and cosmological properties of the absorption systems, we have obtained, in the framework of an ESO key programme devoted to the study of the absorption spectra of high $z$ quasars (D'Odorico et al. 1993), a high resolution spectrum ($R \sim 22000$) of the quasar Q0055-269 at $z = 3.66$, extending from about 4750 to 6300 Å.

We adopt throughout the value $H_o = 50$ km s$^{-1}$ Mpc$^{-1}$ for the Hubble parameter and $q_o = 0.5$.

## 2. DATA ACQUISITION AND REDUCTION

The quasar Q0055-269 has been observed at ESO (La Silla) in Aug 1991, Nov 1991 and Oct 1992, with the NTT telescope and the EMMI instrument in the echelle mode (see D'Odorico 1990). The journal of observations is reported in Table 1.

All the spectra were obtained at low air masses (below 1.3). Particular attention was paid in order to minimize the effects of the atmospheric dispersion by aligning the slit along the average direction of the parallactic angle.

The absolute flux calibration was carried out by observing the standard stars EG274 (Stone & Baldwin 1983), Feige 110 (Stone 1977, Massey & Strobel 1988), Hiltner 600 (Stone 1977, Massey & Strobel 1988). The data reduction has been performed using the standard echelle package described in the 92NOV edition of the MIDAS software (Banse et al. 1988). Wavelengths have been corrected to vacuum heliocentric values and fluxes have been dereddened for galactic extinction according to the Savage and Mathis (1979) curves. A value of $E_{B-V} = 0.016$ has been assumed on the basis of the Burstein and Heiles maps (1982). The weighted mean of the spectra has been obtained at the resolution $R = 22000$ derived from the measure of the sky emission lines. The signal-to-noise ratio per resolution element at the continuum level ranges from $S/N = 12$ to 40 in the interval 4750–6300 Å.

In addition to the EMMI data, two low resolution (R = 100 – 200) HST spectra of Q0055-269 have been retrieved from the HST archive. The two exposures, of 20 and 10 min, were obtained on Sep 10, 1992 with the Faint Object Spectrograph, using a square slit of 4.3 arcsec.

The spectral ranges were $\lambda\lambda = 1200 - 2500$ and $\lambda\lambda = 2000 - 6000$ Å, respectively.

## 3. THE DETECTION AND MEASURE OF ABSORPTION LINES

As usual, a local continuum has been determined selecting regions free of strong absorption lines, where the r.m.s. fluctuation about the mean becomes consistent with noise statistics (see Young et al. 1979). A smoothed continuum has been obtained using spline fits to these spectral regions. The spectrum has then been normalized to the continuum and the observed portion of the Lyman−$\alpha$ forest is shown in Fig. 1 while Fig. 2a,b shows the spectrum longward of the Lyman−$\alpha$ emission.

The detection and measure of absorption lines in the spectrum has been carried out as in GCFT and we refer to this paper for details of the procedure. Here we briefly summarize the main steps. A Gaussian profile is fitted to the high intensity side of the histogram of the pixel intensities of the normalized continuum. The variance obtained is taken as a (conservative) estimate of the noise level. All the lines whose central relative intensity is $3\sigma$ below the continuum are selected to form a statistically well defined sample. The selection method used here corresponds to a well defined locus in the $b - \log N_{HI}$ plane, as shown in Fig. 4.

A $\chi^2$ fitting of Voigt profiles, convolved with the instrumental spread function, was adopted to derive the redshift $z$, the Doppler parameter $b = \sqrt{2}\sigma$ and the column density $N$ for isolated lines and individual components of the blends. The fit was performed with the FITLYMAN program which is an improved version of the code used in our previous work (GCFT) now introduced in the MIDAS package by one of us (A.F.). The number of components of each absorption feature is assumed to be the minimum giving a probability of random deviation $P > 0.05$. The list of all identified absorption lines is given in Table 2.

Heavy element systems have been identified using the atomic parameters given by Morton (1991) and considering a subset of the lines most frequently seen in QSO absorption spectra. Redshift systems have been selected adopting the method of Young et al. (1979), i.e. searching for significant excesses of identifications at all possible redshifts in the wavelength range observed.

The two low resolution HST spectra of Q0055-269 have been used to derive upper limits to the hydrogen column densities of some identified metal line systems. The $\lambda\lambda = 2000 - 6000$ one is shown if Fig. 3. In the $\lambda\lambda = 1200 - 2500$ one no flux below 2300 Å is detected due to a Lyman edge at this wavelength. When the Lyman edge is absent, an upper limit $\log N_{HI} = 16 - 16.5$ has been estimated for the correspondent system from the noise level. A similar analysis has been done on optical data by Sargent, Steidel & Boksenberg (1989).

Nine metal systems have been identified. Furthermore, a possible MgII doublet has been found in the Lyman−α forest. Steidel (1990) identified also two CIV doublets ($z = 3.6013$ and a $z = 3.6763$) which lie outside our wavelength range. For the former system we have fitted the relevant Lyman−α complex. The latter, already classified as uncertain by Steidel, doesn't show any corresponding Lyman−α line, and has to be considered as a data artefact.

We have used a photoionization code (CLOUDY, Ferland 1991) to derive some constraint on the ionization state of the metal systems.

A brief summary for each system is given below:

**z= 0.80411; 1 component** This is a possible weak MgII doublet, since both lines fall in the Lyman−α forest. The corresponding Lyman−α line is not observable due to the presence of a higher-$z$ LLS ($z = 1.53345$).

**z= 1.26765; 2 components** This system is identified by a clear MgII doublet, which exhibits two components. In addition MgI is seen, while no Fe II is detected. HST spectra show an absorption feature in correspondence of the expected position of the Lyman−α line. From the line profile we have estimated an hydrogen column density in the range $17.5 < N_{HI} < 19$. The relevant Lyman Limit is unobservable due to the presence of a higher-$z$ LLS ($z = 1.53345$).

**z= 1.53345; 1 component** This system is responsible for the Lyman Limit trough seen in the HST spectra. The Lyman−α absorption line is also clearly visible. The main features are the FeII transitions (FeII(2344), FeII(2374), FeII(2382), FeII(2586), FeII(2600)) which lie in the red part of the spectrum and have been fitted simultaneously. The ZnI(2139) and ZnII(2062) lines have tentatively been identified in the Lyman $\alpha$ forest. The column density of $\log N_{ZnI} = 12.4$ and $\log N_{ZnII} = 12.7$ for the fitted lines should be considered as upper limits. The value of the hydrogen column density, as measured by the jump at the Lyman edge, is $\log N_{HI} = 17.7$. This value seems too low when compared with the width of the Lyman−α line, from which a value of $19 < \log N_{HI} < 20$ can be inferred. To derive a lower limit to the metallicity of this system we use a reference value $\log N_{HI} = 20$. If the Zn identification is correct, this metal system represent one of the lowest absorption redshifts for which zinc has been found. In a recent paper on metal determinations from ZnII measurements, Pettini et al. (1994) discuss metal enrichment in a sample containing 1/3 of the Damped Lyman $\alpha$ systems known. While they argued that at $z \simeq 2$, metallicity can be reasonably estimated to be around 1/10 of solar, an increase of metallicity with decreasing redshift is considered a premature conclusion. Other observations at lower redshift ($z \leq 1$) are necessary to settle this issue. In this metal system, the metallicity of $[M/H] = -1.8$, found from the FeII column density, is lower than any value measured by Pettini et al. For the adopted value of $N_{HI}$ and $N_{ZnII}$, we have $[M/H] \sim 0$. The discrepancy can be probably attributed to a contamination of Zn lines by Lyman$-\alpha$, to uncertainty in the $N_{HI}$ value and to depletion of the FeII lines by dust grains.

**z= 2.79537; 1 component** This is a low ionization system where SiII(1526), SiII(1260) and CII(1334) are present. An upper limit $\log N_{HI} \leq 16.5$ is derived from the absence of the Lyman edge in HST and optical spectra. From the photoionization code CLOUDY we derive a good fit with a value for the dimensionless ratio of photon to hydrogen densities $\Gamma$ (Ferland, 1991) of $\log \Gamma = -3.1$ and a lower limit to the metal abundance $[M/H] \geq -1$.

**z= 2.87357; 1 component** This system is identified by SiII(1526) and SiII(1260), that has been fitted simultaneously. CII(1334), if present, is hidden in a broad Lyman$-\alpha$ complex.

**z= 2.95044; 2 components** The system is identified in the red part of the spectrum by the CIV doublet, the SiII(1526) and the AlII(1670) (only the first component). CII(1334) and SiIII(1206) are also identified in the Lyman$-\alpha$ forest. With $\log N_{HI} = 16$ we found $\log \Gamma = -3$ and a lower limit $[M/H] \geq -0.6$ in Carbon. Silicon appears 3-5 times overabundant.

**z= 3.085687; 2 components** Identified on the basis of a two-component CIV doublet. No SiIV or low-ionization lines.

**z=3.19 ; 4 components** This is the system extending on the largest velocity range, the total width being approximately 300 km/s. It is made of high ionization lines (CIV and SiIV doublets), showing four components. The third component of the CIV has not been fitted, due to a noisy pattern in the spectrum. Moreover, a single component SiIII(1206) has been found. From the first two CIV and SiIV components, assuming $\log N_{HI} = 16$, we derive $\log \Gamma = -2.5$ and $[M/H] \geq -0.7$.

**z= 3.2558; 1 component** This is a CIV doublet, with no SiIV or low-ionization lines.

**z= 3.43678; 2 components** Identified on the basis of a two-component CIV doublet. No SiIV or low-ionization lines.

In summary, abundances of the 4 metallic systems for which an estimate of the N(HI) was possible, indicate a metal content in the range $[M/H] -1.8/-0.6$. High resolution HST spectra with good S/N are needed to improve these estimates.

## 4. LYMAN ALPHA STATISTICS

### 4.1 Doppler and Column Density Distributions as a Function of z

A statistically well defined sample of Lyman−α lines in the region between Lyman−α and Ly−β emissions ($\lambda\lambda 4810 - 5658$), can be obtained from Table 2. The distribution of the lines in the $b - \log N_{HI}$ plane is shown in Fig. 4 and the dotted curve of constant central line flux represents our selection criterion: only lines with central flux less than the threshold are included in the sample. All the lines with $b < 8$ km s$^{-1}$ have also been removed from the sample since the uncertainties in the fits are large. Besides, some of these lines could be spurious or unidentified metal systems.

From Fig. 4 it appears that saturated lines are not uniformly distributed in the same range of $b$ occupied by unsaturated lines. In particular, there is a deficit of lines with $b < 20$ km s$^{-1}$ and $\log N_{HI} > 13.5 - 14$. Moreover, the strongest lines are in blended features which are difficult to resolve. Such a behaviour has been noted by Carswell et al. (1984) in the spectrum of 1101-264 ($z = 2.14$) and confirmed by the observations of PKS 2126-158 at $z = 3.27$ (GCFT) and by the present sample at $z = 3.66$.

In Fig. 5 the Doppler distribution of the lines not affected by the proximity effect ($8h_{100}$ Mpc, see Bajtlik, Duncan & Ostriker 1988; Lu et al. 1991) is shown. The peak value, $b = 23$ km s$^{-1}$, corresponds, assuming thermal broadening, to $T \simeq 32000$ K. Lines with $b \lesssim 20$ km s$^{-1}$, if real, could provide important information on the physical structure of the clouds. In our Lyman−α sample, 70 lines, i.e. $\simeq 25\%$, have $10 \leq b \leq 20$, and 37, i.e. $\simeq 13\%$ have $15 \leq b \leq 20$, independently of their column densities. The fraction of narrow lines is comparable with the one found in PKS 2126-158.

On the other hand, we are aware that the observed fraction of narrow lines depends on the signal-to-noise ratio of the spectrum. The most extensive study in this respect has been done by Rauch et al. (1993) who in particular showed how the existence of very narrow ($b < 15$ km s$^{-1}$) lines is questionable at low s/n. However, from their simulated spectra it appears that for s/n $\gtrsim 8$, typical of most of our wavelength range, the expected effect on the median of the distribution of the fitted $b$ values is weak. Thus, although an estimate of the intrinsic fraction of lines with $15 < b \lesssim 20$ km s$^{-1}$ would require data at considerably higher s/n, the presence of such lines suggests interesting solutions for the photoionization models of the Lyα clouds.

The column density distribution of the Lyman−α lines is shown in Fig.6. For values $\log N_{HI} \lesssim 13.3$ a selection bias is expected since (see Fig. 4) only the narrower lines are included in the sample. This is confirmed by the drop in the column density distribution below $\log N_{HI} = 13$. A value $\log N_{HI} = 13.3$ has been adopted as completeness limit of the sample (see the dashed line in Fig. 4).

Performing a maximum likelihood analysis of the $N_{HI}$ distribution in the redshift range $z = 2.957 - 3.535$, with the usual assumption of a power-law shape $N_{HI}^{-\beta}$, a value $\beta = 1.8$ is found (Table 3), consistent with the values reported by similar previous analyses (see e.g. Rauch et al. 1992).

However, the Kolmogorov probability for the single power-law shape is only $P_{KS} = 7 \times 10^{-4}$, with a strong indication of a break or cutoff in the column density distribution of Lyman$-\alpha$ clouds. A similar result has been obtained by GCFT for the PKS2126-158 sample but with marginal significance due to the smaller number of lines. A flattening in the column density distribution has been suggested also by Petitjean et al. (1993b) and Meiksin & Madau (1993) using lower resolution data ($\sim 30$ km/s) collected in the literature. Above $\log N_{HI} \sim 14$ a formal value of $\beta \simeq 2.4$ is obtained, but is also possible that these lines are systematically unresolved blends and a real cutoff is present in the intrinsic column density distribution of the optically thin Lyman$-\alpha$ lines. With the present resolution and $S/N$ it is not possible to discriminate between the two options. Removing from the sample few lines with $\log N_{HI} \gtrsim 14.5 - 14.8$ reduces the power-law index from 1.8 to $1.4 - 1.5$, respectively (Fig. 6).

To clarify this point, we have merged the present data with the sample described in GFCT and with the data of 2206-199 (Rauch et al. 1993). Assuming an evolution of the line distribution per unit column density

$$\rho \equiv \frac{\partial^2 n}{\partial z \partial N_{HI}} \propto (1+z)^\gamma N_{HI}^{-\beta} \qquad (1)$$

with $\beta$ independent of redshift, we have carried out a Maximum Likelihood analysis whose results are shown in Table 3.

Considering all the lines with $\log N_{HI} \geq 13.3$ and excluding the region affected by the proximity effect, the best fit values are $\gamma = 2.2$ and $\beta = 1.8$, but the single power-law column density distribution is rejected with a Kolmogorov probability $P(\beta) = 2 \times 10^{-5}$. Excluding lines with $\log N_{HI} > 14.5$ the best fit values become $\gamma = 2.5$ and $\beta = 1.4$, with an acceptance probability $P(\beta) = 0.3$ for the $N_{HI}$ distribution. The power-law distribution in $(1+z)$ is always accepted at $\sim 2\sigma$ level.

It is to note that a steeper value $\beta \sim 2$ is found if we limit the sample to lines with $13.3 \leq \log N_{HI} \leq 14.8$ and $b \leq 20$ km s$^{-1}$. This trend confirms the deficit in the overall sample of strong lines with low $b$ values.

*4.2 Ionization state of the Lyman$-\alpha$ clouds*

Specific features in the intrinsic column density distribution have relevant consequences on the estimate of the general ionizing UV flux (see Sect. 5) and of the physical properties of the clouds. At $N_{HI} \gtrsim 10^{14.5}$ cm$^{-2}$ gravitation could control the dynamics of the clouds and generate a signature in the $N_{HI}$ distribution.

Giallongo & Petitjean (1994) have shown that, assuming photoionization and thermal equilibrium, clouds could have total column densities $\log n_H \simeq -4, -4.3$ and temperatures $T = 20000 - 45000$ K.

We have compared this model with the present data sample at $\langle z \rangle = 3.3$. We have averaged the observed column densities and Doppler parameters, i.e. temperatures, in five bins (Fig. 7a), limiting the sample to lines with $10 \leq b \leq 40$ km s$^{-1}$ to avoid any possible contamination by metal lines and poorly deblended complexes. The average values obtained are not representative of the intrinsic properties of the clouds because of selection effects favoring the detection of narrower lines for $\log N_{HI} < 13.3$. Nonetheless they correspond to a range of physical conditions that should be explained by the model.

As shown by Giallongo & Petitjean (1994), the different temperatures shown in Fig. 7a can be reproduced assuming different amplitudes of the break at 4 Ryd, just beyond the HeII edge, in the range $B = J(4Ryd)/J(4.5Ryd) \sim 1 - 10^3$. In particular, temperatures $\gtrsim 40000$ K imply a featureless power-law ionizing flux, values $\sim 35000$ K with $B = 10$, and values $\sim 19000$ K require $B = 10^3$.

In this model, lines with $\log N_{HI} > 13.5$ and $T > 35000$ K correspond to clouds with sizes $S > 50$ kpc (Fig. 7b), in agreement with the lower limits obtained from the gravitationally lensed quasar pairs.

For the same ionization and thermal states, the Jeans length ranges from 350 to 500 kpc and the column densities beyond which the clouds are unstable are plotted in Fig. 4 and Fig. 7a as a continuous curve. The column density dependent cutoff observed in our metal-free Lyman$-\alpha$ sample then arises naturally.

### 4.3 Clustering properties of the Lyman$-\alpha$ clouds

No clustering in the velocity space has been detected so far for Lyman$-\alpha$ lines on scales $300 < \Delta v < 30000$ km s$^{-1}$ (Sargent et al. 1980, 1982; Bechtold 1987; Webb & Barcons 1991). Their distribution is different from that of metal-line systems selected by means of the CIV doublet (Sargent et a. 1988), where strong clustering is present up to 600 km s$^{-1}$.

Preliminary results at higher resolution seem to indicate weak clustering on smaller scales ($\Delta v = 50 - 300$ km s$^{-1}$, Webb 1987). However up to now no signal, even at the smallest observable separation, has been detected in individual high resolution spectra (Pettini et al. 1990, Rauch et al. 1992) even if Rauch et al. have claimed low doppler parameter lines to be clustered, but not the broad lines.

In the present case better chances are offered by the larger density of lines observed at the relatively higher redshift of our sample. We follow earlier work in considering the two-point correlation function, defined as the excess, due to clustering, of the probability $dP$ of finding a Lyman$-\alpha$ cloud in a volume $dV$ at a distance $r$ from another cloud

$$dP = \Phi_{Ly\alpha}(z)dV[1+\xi(r)] \qquad (2)$$

where $\Phi(z)$ is the average space density of the clouds as a function of $z$. Since we want to study clustering on small velocity scales, then the cosmological change in the distance scale with cosmic time can be neglected and the velocity difference can be simply deduced from the redshift difference (Sargent et al. 1980)

$$\Delta v = \frac{c(z_2 - z_1)}{1 + (z_1 + z_2)/2} \qquad (3)$$

where $\Delta v$ is the velocity of one cloud as measured by an observer in the restframe of the other.

The two-point correlation function in the velocity space is defined as (Peebles 1980)

$$\xi = \frac{N_{obs}(v, \Delta v)}{N_{exp}(v, \Delta v)} - 1 \qquad (4)$$

where $N_{obs}$ is the observed number of line pairs with velocity separations between $v$ and $v + \Delta v$ and $N_{exp}$ is the number of pairs expected from a random distribution in redshift.

In our line sample $N_{exp}$ is obtained averaging 1000 numerical simulations of the observed number of redshifts randomly generated in the same redshift interval as the data. The set of redshifts is obtained according to the cosmological distribution $\propto (1+z)^\gamma$ where the best value is obtained from the maximum likelihood analysis of the real line sample in the given interval of column density and redshift. The results are not sensitive to the value of $\gamma$ adopted and even a flat distribution (i.e. $\gamma = 0$) gives values of $\xi$ that differ typically by less than 0.02.

With this random generation of lines, it is possible to correct for incomplete wavelength coverage due to gaps in the spectrum or line blanketing of weak lines due to strong complexes. No velocity splittings $\Delta v < 20$ km s$^{-1}$ were included in the estimate of $N_{exp}$ because of intrinsic line blending due to the typical widths of the Lyman$-\alpha$ lines. The $1\sigma$ deviation from a random distribution is given by the $N_{obs}^{-1/2}$ resulting in each bin. This is a good approximation in the case of weak clustering $\xi \lesssim 1$.

The resulting correlation function of Lyman$-\alpha$ lines in Q0055-269 with $N_{HI} > 13.3$ is plotted in Fig. 8. It is clear that a weak but significant signal is present with $\xi \simeq 0.34 \pm 0.06$ up to 350 km s$^{-1}$. We have also explored the variations of the correlation function as a function of the column density. It appears that the correlation function increases in the first bin as we increase the minimum column density in the sample. The maximum significant signal is obtained for lines with $\log N_{HI} \gtrsim 13.8$ for which the correlation function increases up to $\xi = 0.89 \pm 0.18$ for $\Delta v = 100$ km s$^{-1}$ (Fig. 8). When computing $\xi$ for lines in the interval $\log N_{HI} = 13.3 - 13.6$ (including in the simulated spectra wavelength gaps due to lines with $\log N_{HI} > 13.8$), no significant clustering is observed. This suggests that the weak clustering observed in the overall sample by Webb (1987) can be due to the larger number of weak lines which are randomly distributed in redshift. Only in QSO spectra at high redshifts, as in our case, the density of lines with $\log N_{HI} > 13.8$ is large enough to show significant clustering.

For comparison we have computed the correlation function of the Lyman−$\alpha$ samples obtained from the quasar PKS 2126-158 (GCFT) and 0014+81 (Rauch et al. 1992). A significant clustering appears in Fig. 9 when the sample is limited to strong lines with $\log N_{HI} \gtrsim 14$ even if the significance depends on the number of absorption lines in the individual spectra. Indeed, we have obtained $\xi = 1.02 \pm 0.26$ for $\Delta v = 100$ km s$^{-1}$ for lines with $\log N_{HI} \geq 13.8$ in PKS 2126-158 and $\xi = 0.85 \pm 0.30$ for lines with $\log N_{HI} \geq 14.1$ in 0014+81.

A similar claim has been done by Crotts (1989) on the basis of a cross-correlation analysis of lines in the two spectra of quasar pairs. However, the low resolution of the spectra didn't allow the detection of line correlation in individual spectra.

The correlation found in the Lyman−$\alpha$ cloud positions is less pronounced than that observed in metal-line absorption systems (Sargent et al. 1988) or galaxies (Peebles 1980), but consistent with a scenario of gravitationally induced correlations, as expected in models where gravitation is an important confining agent of the Lyman−$\alpha$ clouds. The correlation found would imply, in the standard CDM framework, that the Lyman−$\alpha$ clouds of larger column density can trace the mass while the clouds responsible for the weakest absorption could be abundant in underdense regions. In the latter case the redshift autocorrelation function could be lower than their spatial autocorrelation (Kaiser 1987; Mo, Miralda-Escudé, & Rees 1993).

In any case, deriving the three-dimensional spatial autocorrelation function of the clouds, $\xi_r$, from the redshift correlation, $\xi_v$, is not an easy task, as emphasized by Heisler, Hogan, & White (1989). At small velocity differences the two functions are related by the simple convolution (Heisler et al. 1989):

$$\xi_v = \int_0^\infty H dr \; \xi(r) \; P(v \mid r) \propto \int_{r_{cl}}^\infty \frac{H dr}{\sigma} \left(\frac{r}{r_o}\right)^{-\gamma} \times \left\{ exp\left[-\frac{(Hr-v)^2}{2\sigma^2}\right] + exp\left[-\frac{(Hr+v)^2}{2\sigma^2}\right] \right\} \tag{5}$$

where a Gaussian distribution of the peculiar motions respect to the Hubble flow has been assumed together with a power-law spatial correlation function of the galaxy-type.

At small velocity splittings, the redshift correlation scale $r_o$ depends mainly on the cloud sizes $r_{cl}$ and on the velocity dispersion assumed. Although these quantities are very poorly known, it is interesting, with this simple approach, to try to derive constraints on the cloud sizes and velocity dispersions from the observed correlation.

Assuming an index $\gamma = 1.77$ for the power-law spatial correlation function (as commonly measured for galaxies), the fit shown in Fig. 8 results in an upper limit on the cloud sizes as a function of the velocity dispersion and of the appropriate correlation scale. We obtain $r_{cl} \sim 100 - 190$ kpc for $\sigma_v = 50 - 160$ km s$^{-1}$, and $r_o \sim 245 - 420$ kpc, respectively. In the former case, however, a good fit to the overall shape of the observed correlation function would require a very low value $\gamma = 1.1$. For $\sigma_v > 200$ km s$^{-1}$ the redshift correlation function becomes too flat and very steep $\gamma$ values are needed to find an acceptable fit to the data (e.g. $\sigma_v = 300$ km s$^{-1}$ requires $\gamma = 4$). In this scenario the cloud sizes would then be confined in the interval $r_{cl} \sim 30 - 200$ kpc, where the lower limit is constrained from the observations of coincident absorptions in quasar pairs.

The upper limits on the sizes obtained from this clustering analysis are consistent with the values of $r_{cl} \sim 100 - 180$ kpc derived from the photoionization model described in Giallongo & Petitjean (1994) and in Sect. 4.2, applied to Lyman$-\alpha$ lines with $\log N_{HI} \gtrsim 13.8$.

The aim of this simple exercise is to emphasize the importance of detecting a clear signal and a detailed shape in the correlation function of the strong Lyman$-\alpha$ lines at high redshifts, for which a larger sample is needed.

### 4.4 Voids in the Lyman$-\alpha$ forest

Several authors have searched for megaparsec-sized voids in the Lyman$-\alpha$ forests of quasars. Two previous claims exist: one for Q0420-388, a void with comoving size $\sim 30$ Mpc, whose statistical significance has been questioned (Crotts 1987,1989; Ostriker, Bajtlik, & Duncan 1988); another one in Q0302-003, with a size $\sim 23$ Mpc, detected by Dobrzycki & Bechtold (1991).

Voids in the Lyman$-\alpha$ forest can be a useful test for theoretical models concerning the large-scale structure in the Universe, even if changes in the ionization of the IGM due to fluctuations in the ionizing UV flux can be efficient in depleting neutral hydrogen along the line of sight. Indeed, Dobrzycki & Bechtold (1991) suggested that the void in Q0302-003 can result from enhanced ionization by the nearby foreground quasar Q0301-005.

In any case, the statistical significance of a given void depends exponentially on the line density (Ostriker et al. 1988) and uncertainties in the line statistics strongly influence the probability estimate of the void. High resolution data which are less affected by blending effects are ideal in this respect. We have looked for gaps in our spectrum by analysing the subset of lines with $\log N_{HI} \geq 13.3$ and $b \lesssim 30$ km s$^{-1}$, to ensure statistical completeness and avoid deblending uncertainties. We have found two regions devoid of such lines centered respectively at $\lambda \sim 5000, 5206$ Å, with sizes $\Delta v \sim 2009, 2046$ km s$^{-1}$, corresponding to a $\sim 20$ Mpc size and a $\sim 118$ Mpc separation.

To establish the random probability of observing such gaps, a set of 50000 spectra have been simulated with the observed number ($N = 178$) of redshifts randomly generated in the same redshift interval as the data. The region within 8 Mpc from the quasar has been avoided because of the proximity effect. The set of redshifts is obtained according to the cosmological distribution $A(1+z)^\gamma$, where the best value is obtained from the maximum likelihood analysis of the real line sample in the given interval of column density and redshift. A best fit value of $\gamma = 1.5$ has been found from the Lyman$-\alpha$ sample of Q0055-269 alone.

We have obtained 932 gaps larger than 2046 km s$^{-1}$. The random probability to get a gap larger than the maximum observed gap is then $\simeq 0.02$. The results are not sensitive to the value of $\gamma$ adopted, in agreement with Dobrzycki & Bechtold (1991), even a flat distribution (i.e. $\gamma = 0$) gives the same probability.

The 2046 km/s gap is significant only to $2.2\sigma$ level, but the joint probability for the presence of the two observed gaps in the same spectrum is only $\simeq 2 \times 10^{-4}$.

Even if underdense regions are statistically significant in our spectrum, the filling factor is less than 10%, in agreement with previous estimates (Ostriker, et al. 1988, Carswell & Rees 1987, Crotts 1987).

## 5. THE PROXIMITY EFFECT

The proximity effect consists in a reduction of the line density in the region affected by the quasar ionizing flux. When the factorization $\rho = (1+z)^\gamma f(N_{HI})$ is possible, it is convenient to define the coevolving redshift interval $dX_\gamma = (1+z)^\gamma dz$. This allows the comparison of the $N_{HI}$ distributions $\partial^2 n/\partial X_\gamma \partial N_{HI} = f(N_{HI})$, independently of redshifts.

If the clouds are highly ionized by the UV background and optically thin, the neutral hydrogen column density $N$ of a cloud near a quasar is (the subscript HI is omitted for simplicity):

$$N = \frac{N_\infty}{1+\omega} \qquad (6)$$

where $N_\infty$ is the *intrinsic* column density the same cloud would have at infinite distance from the quasar and $\omega(z, z_Q) = F_\nu/4\pi J$ is the ratio between the Lyman limit QSO flux impinging on the cloud and the background flux $4\pi J$ at the same frequency (see Bajtlik et al. 1988).

Assuming a power law distribution in a given interval $(N_{min}, N_{\infty,max})$, then the number of lines in the coevolving redshift interval with a given $\omega$ is (GCFT):

$$\frac{\partial n}{\partial X_\gamma}(\omega) = \frac{N_{\infty,max}^{1-\beta} - N_{min}^{1-\beta}(1+\omega)^{1-\beta}}{1-\beta}, \tag{7}$$

where the lower limit $N_{min}$ can be fixed in correspondence to the selection threshold adopted to ensure a reasonable completeness of the line sample. Thus, the function becomes zero when the local ionization shifts the upper $N$ cut-off down to the adopted lower threshold.

For the 3 quasars used in our previous analysis, we have adopted the unbiased quasar redshifts and the Lyman limit fluxes as in GCFT. For 2206-199 and 0055-269 we have adopted the values $z = 2.56, 3.66$ for the redshifts and $F_\nu = 2.4, 1.5$ (in units of $10^{-27}$ ergs s$^{-1}$ cm$^{-2}$ Hz$^{-1}$), respectively.

In Fig. 10, $\partial n/\partial X_\gamma$ is shown as a function of $\omega$, together with the data points binned as in Bajtlik et al (1988) for given values of $J$. Assuming that $J$ does not change in the redshift interval covered by the data, we consider lines in the interval $13.3 \leq \log N_{HI} \leq 14.2$, to avoid strong deviations from the single power-law and contaminations by the strong lines present near the emission redshift of the quasar. These lines could indeed be due to material ejected by the quasar.

A good fit is obtained assuming the value $5 \times 10^{-22}$ $cgs$, which is not far from the value derived from the QSO contribution ($3 \times 10^{-22} cgs$) in the same $z$ interval (Meiksin & Madau 1993). Taking into account slope uncertainties of the faint-end QSO luminosity function and uncertainties in the low-luminosity cutoff, the estimated QSO contribution to the ionizing UVB could be a lower limit and should be considered consistent with the value derived from the proximity effect.

Estimates of the redshift dependence of the UVB await for a larger line sample at high resolution. Nonetheless, we find a good agreement between theory and observations mainly due to the improved values $\gamma$ and $\beta$ of the column density distribution derived from a high resolution sample, and therefore less affected by blending.

## 6. CONCLUSIONS

A sample of absorption lines has been extracted from the spectrum of Q0055-269 ($z_{em} = 3.66$) taken at the resolution of 14 km s$^{-1}$ ($b \simeq 8$ km s$^{-1}$) over the range $4750 < \lambda < 6300$ Å. The main results are summarized as follows:

- Nine metal systems have been identified. The absence of Lyman Limit absorption in most of the heavy element systems found in Q0055-269 seems to favour high metallicities

- $[M/H] \gtrsim -1$. High resolution HST spectra with good S/N are needed to improve these estimates.

- The sample of Lyman$-\alpha$ lines not associated with identified metal systems has a Doppler parameter distribution peaked at $b \simeq 23$ km s$^{-1}$ with 13% of lines having $15 \leq b \leq 20$ km s$^{-1}$, independently of their column densities. It is not possible to establish the reality of lines with $b < 15$ km s$^{-1}$ given the signal-to-noise ratio present in our spectrum. Besides, a deficit of isolated lines with $b \lesssim 20$ km s$^{-1}$ and $\log N_{HI} > 13.5$ is observed.

- A single power-law distribution represents a poor fit to the column density distribution. Merging the present sample with other high resolution data available in the literature, a Kolmogorov probability $P_{KS} = 10^{-5}$ is obtained. The reality of stronger lines is doubtful and in any case their distribution appears steeper beyond $\log N_{HI} \sim 14$, with an index $\beta \sim 2$. A better fit is given by a flatter power-law ($\beta \simeq 1.4$) in the range $13.3 \lesssim \log N_{HI} \lesssim 14.5$ with a cutoff or a steepening at larger column densities.

- Both the presence of a cutoff in the column density distribution and of temperatures as low as 20000 K (i.e. $b \simeq 18$ km s$^{-1}$) are found to be consistent with a photoionization model where clouds have low total densities $n_H \sim 10^{-4}$ cm$^{-3}$ and large sizes $S > 20$ kpc.

- Clustering in the redshift space has been found for the Lyman$-\alpha$ lines. The amplitude of the clustering increases when increasing the column density of the lines. In particular $\xi$ is about 1 up to $\Delta v = 150$ km s$^{-1}$ for lines with $\log N_{HI} \gtrsim 13.8$. A simple model, in which spatial clustering of the type observed for galaxies is convolved with gaussian peculiar motions, gives upper limits to the cloud sizes $S \sim 200$ kpc.

- Two voids with size $\Delta v \simeq 2000$ km s$^{-1}$, where no lines with $\log N_{HI} > 13.3$ and $b < 30$ km s$^{-1}$ are present, have been found in the spectrum of Q0055-269. The joint random probability is only $\simeq 2 \times 10^{-4}$.

- The effect of breaks or cutoffs has been included in the analysis of the proximity effect. A value for the ultraviolet background flux $J = 5 \times 10^{-22}$ ergs s$^{-1}$ cm$^{-2}$ Hz$^{-1}$ sr$^{-1}$ has been found in the redshift interval $1.8 \lesssim z \lesssim 3.66$. This value is consistent with the integrated QSO contribution to the UVB.

**ACKNOWLEDGMENTS** We thank P. Petitjean for discussions. E.G. acknowledges the support of ESO as a visiting astronomer while part of this work was done. This research is partly supported by an EC HCM programme.

# FIGURE CAPTIONS

**Fig. 1.** Echelle spectrum of Q0056-269 in the Lyman alpha forest normalized to unit continuum. Upper and lower marks show Lyman$-\alpha$ and heavy element components, respectively, listed in Table 2. The long upper marks indicate Lyman$-\alpha$ lines with $10 \leq b \leq 20$ km s$^{-1}$.

**Fig. 2.** Normalized Echelle spectrum of Q0056-269 redward of the QSO Lyman$-\alpha$ emission from 5700 (top-left) to 6600 (down-right). Thicks on the vertical axis show the zero flux levels and are spaced of 1.5 units.

**Fig. 3.** Low-resolution HST FOS spectrum of Q0056-269. The dashed line indicates the $1\sigma$ noise level.

**Fig. 4.** Plot of the Doppler parameter $b$ vs. $\log N_{HI}$ for the Lyman$-\alpha$ lines listed in Table 2. The dashed curve shows the selection criteria adopted. The continuous line represents the Jeans instability threshold for the model described in Sect. 4.2.

**Fig. 5.** Doppler parameter distribution for the Lyman$-\alpha$ sample in Q0055-269.

**Fig. 6.** Column density distribution of Lyman$-\alpha$ lines out of 8 Mpc from Q0055-269.

**Fig. 7.** a) Temperature vs. HI column density for the Lyman$-\alpha$ lines in Q0055-269. The data have been arbitrarily binned. b) Size vs. HI column density of cloud models with temperatures as in a) and with different shapes of the UV ionizing flux as described in the text.

**Fig. 8.** Two-point correlation function for the Lyman$-\alpha$ lines in Q0055-269. The continuous histogram is for lines with $\log N_{HI} \gtrsim 13.8$; the dotted histogram is for lines with $\log N_{HI} \gtrsim 13.3$. The continuous curve is the model described in Sec. 4.3 and computed from the eq. (5) with $\gamma = 1.77$, $\sigma = 150$ km s$^{-1}$, $r_{cl} = 110$ kpc and $r_o = 280$ kpc at $z = 3.3$.

**Fig. 9.** Two-point correlation function for the Lyman$-\alpha$ lines with $\log N_{HI} \gtrsim 14$. Dotted histogram is for the quasar Q0014+81, continuous histogram for the quasar PKS 2126-158.

**Fig. 10.** $\partial n / \partial X_\gamma$ as a function of $\log \omega$ for Lyman$-\alpha$ lines in Q0055-269 in the interval $\log N_{HI} = 13.3 - 14.2$ Filled circles correspond to $J_{-22} = 5$.

TABLE 1
Journal of Observations

| Date (yyyy mm dd) | Integration time (s) | Seeing | Slit | Wavelength range (Å) | CCD |
|---|---|---|---|---|---|
| 1991 08 06 | 9127 | 1".0 | 1".5 | 4148–7019 | FORD 2k |
| 1991 11 06 | 8895 | 1".1 | 1".2 | 4253–6600 | Th 1k ♯18 |
| 1991 11 07 | 7200 | 1".1 | 1".2 | 4253–6600 | Th 1k ♯18 |
| 1992 10 19 | 6000 | 0".9 | 1".2 | 4384–6600 | Th 1k ♯18 |
| 1992 10 21 | 6000 | 1".0 | 1".4 | 4415–6600 | Th 1k ♯18 |

# TABLE 2
# Q0055-269 Absorption Line Parameters

| $\lambda_{vac}$ | ± | $b$ | ± | $\log N$ | ± | ID | $\lambda_{vac}$ | ± | $b$ | ± | $\log N$ | ± | ID |
|---|---|---|---|---|---|---|---|---|---|---|---|---|---|
| 4766.14 | .02 | 18.10 | 2.20 | 13.44 | .06 | SiIII 1206 $z$=2.95 | 5013.21 | .03 | 10.00 | 3.80 | 13.04 | .08 | |
| 4766.82 | .01 | 6.20 | 2.70 | 14.53 | 1.58 | SiIII 1206 $z$=2.95 | 5016.67 | .02 | 21.90 | 2.10 | 14.47 | .16 | |
| 4783.65 | .05 | 2.00 | .80 | 12.73 | .33 | SiII 1260 $z$=2.79 | 5019.48 | .02 | 40.10 | 2.60 | 14.31 | .05 | |
| 4800.56 | .13 | 46.90 | 6.70 | 13.98 | .11 | Lya $z$=2.95 | 5022.06 | .08 | 17.60 | 9.80 | 13.04 | .15 | |
| 4802.29 | .12 | 68.00 | 4.30 | 14.56 | .05 | Lya $z$=2.95 | 5023.45 | .18 | 41.20 | 11.40 | 13.91 | .13 | |
| 4804.48 | .06 | 53.00 | 2.80 | 14.29 | .04 | Lya $z$=2.95 | 5024.43 | .03 | 29.60 | 2.70 | 13.93 | .04 | |
| 4812.17 | .04 | 23.10 | 3.80 | 13.53 | .05 | | 5025.70 | .10 | 25.30 | 12.80 | 13.09 | .15 | |
| 4813.43 | .04 | 11.50 | 4.10 | 13.05 | .09 | | 5036.05 | .08 | 32.80 | 7.50 | 13.39 | .08 | |
| 4835.60 | .02 | 20.60 | 1.90 | 13.68 | .04 | | 5038.05 | .03 | 32.70 | 4.20 | 14.92 | .24 | |
| 4839.85 | .02 | 22.30 | 2.10 | 13.77 | .04 | | 5039.40 | .03 | 16.40 | 2.90 | 13.63 | .06 | |
| 4841.05 | .08 | 37.70 | 8.60 | 13.42 | .07 | | 5044.92 | .02 | 2.10 | .30 | 13.43 | .44 | MgII 2796 $z$ = 0.8 |
| 4847.64 | .09 | 44.40 | 11.80 | 13.53 | .08 | | 5047.15 | .02 | 2.00 | .50 | 13.26 | .26 | |
| 4849.81 | .05 | 12.90 | 4.40 | 13.39 | .20 | | 5049.03 | .02 | 14.50 | 2.50 | 13.42 | .05 | |
| 4850.49 | .12 | 41.30 | 8.40 | 13.75 | .09 | | 5050.27 | .02 | 2.50 | .50 | 13.72 | .40 | |
| 4852.63 | .04 | 9.90 | 4.00 | 13.00 | .09 | | 5050.85 | .06 | 20.20 | 9.50 | 13.21 | .12 | |
| 4854.05 | .02 | 23.30 | 1.60 | 13.94 | .05 | | 5055.00 | .02 | 18.60 | 1.90 | 14.05 | .09 | |
| 4857.44 | .07 | 56.00 | 6.50 | 14.12 | .04 | | 5056.31 | .02 | 19.50 | 3.00 | 14.30 | .38 | SiIII 1206 $z$ = 3.19 |
| 4859.29 | .07 | 47.60 | 4.70 | 14.08 | .04 | | 5057.87 | .00 | 2.10 | .30 | 13.43 | .44 | MgII 2803 $z$ = 0.8 |
| 4862.58 | .12 | 54.60 | 11.70 | 13.55 | .08 | | 5058.89 | .02 | 24.80 | 2.10 | 14.21 | .08 | |
| 4868.08 | .03 | 28.10 | 3.20 | 13.92 | .04 | | 5061.07 | .05 | 31.80 | 4.60 | 13.55 | .05 | |
| 4869.32 | .03 | 31.10 | 3.70 | 14.05 | .04 | | 5062.28 | .04 | 9.00 | 3.20 | 12.90 | .10 | |
| 4870.87 | .03 | 10.40 | 3.00 | 13.20 | .07 | | 5065.05 | .03 | 9.70 | 3.30 | 13.53 | .08 | CII 1334 $z$ = 3.19 |
| 4871.64 | .02 | 20.10 | 4.50 | 14.11 | .18 | | 5066.69 | .03 | 11.90 | 3.20 | 13.30 | .08 | |
| 4872.51 | .05 | 19.20 | 3.20 | 13.79 | .06 | | 5067.55 | .02 | 23.00 | 2.60 | 14.20 | .09 | |
| 4873.37 | .04 | 25.50 | 3.30 | 13.84 | .06 | | 5070.82 | .04 | 21.60 | 2.80 | 14.21 | .12 | |
| 4874.90 | .01 | 12.80 | 2.70 | 14.42 | .45 | | 5072.31 | .06 | 52.80 | 4.60 | 14.44 | .05 | |
| 4879.47 | .17 | 59.40 | 11.70 | 13.57 | .09 | | 5076.80 | .04 | 24.40 | 5.00 | 13.45 | .06 | |
| 4881.26 | .08 | 43.80 | 8.70 | 13.73 | .07 | | 5078.02 | .03 | 22.50 | 2.30 | 14.09 | .06 | |
| 4882.34 | .03 | 3.10 | 7.50 | 13.01 | .00 | SiII 1260 $z$=2.87 | 5079.35 | .04 | 41.90 | 3.30 | 14.30 | .04 | |
| 4883.96 | .03 | 11.60 | 3.10 | 13.21 | .07 | | 5080.74 | .03 | 13.00 | 2.60 | 13.05 | .06 | |
| 4885.22 | .05 | 27.80 | 5.10 | 13.40 | .06 | | 5081.37 | .02 | 14.30 | 2.20 | 13.59 | .05 | |
| 4887.07 | .03 | 12.40 | 3.10 | 13.05 | .06 | | 5082.74 | .05 | 32.20 | 4.50 | 13.83 | .05 | |
| 4888.35 | .04 | 24.00 | 2.90 | 13.46 | .05 | | 5083.83 | .04 | 23.60 | 3.00 | 13.80 | .05 | |
| 4891.31 | .02 | 6.80 | 2.90 | 13.15 | .11 | | 5085.46 | .03 | 29.10 | 2.60 | 13.79 | .03 | |
| 4898.94 | .04 | 30.80 | 3.00 | 13.57 | .04 | | 5086.90 | .02 | 24.70 | 2.70 | 14.04 | .05 | |
| 4906.49 | .04 | 33.10 | 3.10 | 14.01 | .04 | | 5087.97 | .14 | 53.50 | 4.40 | 13.79 | .03 | |
| 4907.71 | .02 | 19.30 | 3.60 | 14.74 | .40 | | 5090.12 | .03 | 21.30 | 2.50 | 13.77 | .04 | |
| 4909.20 | .03 | 22.50 | 4.50 | 14.63 | .29 | | 5091.30 | .04 | 42.50 | 4.70 | 13.82 | .03 | |
| 4910.23 | .03 | 15.50 | 3.30 | 13.98 | .15 | | 5092.17 | .02 | 2.50 | .90 | 14.07 | .77 | |
| 4911.77 | .03 | 11.70 | 2.70 | 13.13 | .06 | | 5093.24 | .02 | 21.20 | 3.60 | 14.64 | .33 | Lya $z$ = 3.19 |
| 4913.20 | .05 | 35.10 | 3.40 | 14.29 | .05 | | 5094.54 | .05 | 45.30 | 6.40 | 14.99 | .04 | Lya $z$ = 3.19 |
| 4914.58 | .07 | 35.20 | 4.10 | 13.93 | .06 | | 5095.93 | .02 | 22.90 | 3.70 | 14.49 | .21 | Lya $z$ = 3.19 |
| 4916.70 | .02 | 10.70 | 3.20 | 13.33 | .05 | | 5097.11 | .04 | 21.40 | 3.90 | 13.90 | .10 | Lya $z$ = 3.19 |
| 4917.33 | .03 | 13.20 | 3.00 | 13.64 | .06 | | 5098.56 | .07 | 66.30 | 9.80 | 14.36 | .04 | Lya $z$ = 3.19 |
| 4917.86 | .02 | 9.70 | 2.30 | 13.57 | .09 | | 5103.58 | .08 | 20.00 | 8.30 | 13.04 | .12 | |
| 4923.07 | .03 | 19.60 | 2.80 | 13.31 | .04 | | 5106.49 | .05 | 23.30 | 5.40 | 13.26 | .07 | |
| 4925.88 | .03 | 53.50 | 2.70 | 14.13 | .02 | | 5109.99 | .02 | 46.00 | 1.90 | 13.97 | .02 | |
| 4932.98 | .03 | 69.20 | 3.50 | 14.47 | .03 | | 5112.17 | .06 | 23.00 | 5.60 | 13.26 | .07 | |
| 4937.13 | .03 | 14.50 | 1.90 | 13.17 | .05 | | 5114.79 | .05 | 7.90 | 6.90 | 12.71 | .14 | |
| 4938.56 | .03 | 18.40 | 2.50 | 13.77 | .06 | | 5116.91 | .06 | 16.00 | 5.00 | 12.93 | .10 | |
| 4939.56 | .04 | 33.70 | 2.90 | 14.08 | .04 | | 5118.51 | .05 | 1.50 | 5.40 | 13.21 | 2.09 | |
| 4942.19 | .14 | 1.30 | 3.60 | 13.58 | .84 | OI 1302 $z$=2.79 | 5119.80 | .09 | 28.50 | 8.60 | 13.12 | .10 | |
| 4943.11 | .03 | 22.40 | 2.60 | 13.52 | .04 | | 5122.26 | .07 | 21.50 | 5.20 | 13.59 | .10 | |
| 4944.80 | .05 | 29.10 | 5.00 | 13.43 | .06 | | 5123.23 | .07 | 28.90 | 9.80 | 13.79 | .12 | |
| 4946.08 | .03 | 29.60 | 2.50 | 13.72 | .03 | | 5124.02 | .03 | 13.90 | 2.80 | 14.05 | .19 | |
| 4959.42 | .02 | 2.30 | 1.60 | 13.98 | 1.80 | | 5124.73 | .04 | 15.70 | 4.30 | 13.23 | .07 | |
| 4963.13 | .02 | 9.10 | 2.20 | 13.15 | .06 | | 5125.96 | .05 | 27.90 | 4.50 | 13.40 | .05 | |
| 4964.48 | .02 | 32.50 | 2.40 | 14.60 | .11 | | 5126.85 | .03 | 19.30 | 3.60 | 13.42 | .05 | |
| 4966.56 | .04 | 56.70 | 3.70 | 14.43 | .04 | Lya $z$ = 3.08 | 5128.83 | .02 | 18.20 | 2.10 | 13.89 | .06 | |
| 4968.00 | .03 | 12.20 | 3.40 | 13.46 | .07 | Lya $z$ = 3.08 | 5129.58 | .03 | 8.60 | 3.40 | 13.13 | .08 | |
| 4971.04 | .02 | 36.20 | 2.90 | 14.23 | .05 | | 5130.70 | .04 | 16.30 | 2.70 | 12.95 | .06 | |
| 4974.35 | .03 | 31.60 | 2.90 | 13.71 | .03 | | 5132.58 | .03 | 24.20 | 2.70 | 13.85 | .04 | |
| 4976.09 | .02 | 22.90 | 1.80 | 14.05 | .06 | | 5133.78 | .03 | 26.80 | 2.40 | 14.07 | .04 | |
| 4978.95 | .02 | 33.90 | 2.30 | 14.31 | .06 | | 5134.91 | .07 | 24.20 | 5.30 | 12.95 | .08 | |
| 4983.18 | .09 | 70.50 | 7.20 | 13.71 | .04 | | 5136.05 | .03 | 26.60 | 2.50 | 13.66 | .03 | |

| $\lambda_{vac}$ | ± | $b$ | ± | $\log N$ | ± | ID | $\lambda_{vac}$ | ± | $b$ | ± | $\log N$ | ± | ID |
|---|---|---|---|---|---|---|---|---|---|---|---|---|---|
| 5138.65 | .02 | 1.90 | .50 | 13.36 | .26 | | 5328.78 | .05 | 19.60 | 4.60 | 13.10 | .08 | |
| 5139.10 | .02 | 2.10 | .90 | 13.45 | .53 | | 5330.71 | .09 | 64.50 | 5.70 | 13.83 | .04 | |
| 5145.98 | .05 | 10.90 | 4.30 | 12.83 | .11 | | 5331.25 | .04 | 16.50 | 3.30 | 13.44 | .09 | |
| 5148.30 | .05 | 43.00 | 5.40 | 13.72 | .04 | | 5333.53 | .03 | 32.80 | 2.00 | 13.81 | .03 | |
| 5149.42 | .05 | 26.10 | 5.60 | 13.57 | .07 | | 5334.82 | .03 | 13.20 | 3.30 | 13.05 | .07 | |
| 5150.47 | .03 | 26.00 | 2.80 | 13.93 | .04 | | 5338.38 | .05 | 24.90 | 5.20 | 13.35 | .06 | |
| 5153.39 | .04 | 12.40 | 3.30 | 12.94 | .06 | | 5339.53 | .04 | 32.70 | 3.40 | 13.65 | .03 | |
| 5154.47 | .03 | 15.10 | 3.30 | 13.01 | .06 | | 5341.49 | .04 | 20.20 | 3.60 | 13.33 | .05 | |
| 5159.40 | .08 | 50.40 | 8.60 | 13.59 | .05 | | 5343.03 | .10 | 41.60 | 6.70 | 13.48 | .07 | |
| 5162.13 | .08 | 46.50 | 10.10 | 13.59 | .07 | | 5345.17 | .06 | 38.80 | 5.40 | 13.47 | .05 | |
| 5163.53 | .03 | 34.90 | 5.90 | 14.32 | .08 | | 5347.96 | .06 | 65.60 | 7.30 | 13.90 | .03 | |
| 5165.42 | .07 | 66.40 | 9.50 | 14.51 | .04 | | 5351.04 | .02 | 26.70 | 1.80 | 13.77 | .03 | |
| 5167.18 | .03 | 34.80 | 4.10 | 14.09 | .03 | | 5352.06 | .02 | 1.80 | 1.70 | 13.60 | 1.86 | |
| 5169.38 | .03 | 49.10 | 3.60 | 14.29 | .03 | | 5353.15 | .02 | 10.90 | 2.10 | 13.15 | .05 | |
| 5173.79 | .08 | 65.10 | 4.70 | 14.41 | .04 | Lya $z = 3.25$ | 5358.03 | .11 | 54.90 | 8.20 | 13.59 | .06 | |
| 5175.23 | .07 | 35.90 | 4.50 | 14.14 | .06 | Lya $z = 3.25$ | 5358.81 | .03 | 12.00 | 2.40 | 13.26 | .09 | |
| 5181.63 | .06 | 28.90 | 3.40 | 14.15 | .08 | | 5361.45 | .03 | 9.50 | 2.70 | 12.88 | .08 | |
| 5183.03 | .06 | 25.20 | 4.90 | 15.01 | .39 | | 5363.21 | .02 | 28.40 | 1.60 | 14.07 | .03 | |
| 5184.75 | .02 | 4.80 | 4.10 | 13.30 | .48 | | 5364.37 | .04 | 11.80 | 3.70 | 12.93 | .09 | |
| 5185.33 | .02 | 13.80 | 1.70 | 13.48 | .05 | | 5366.04 | .04 | 16.00 | 4.30 | 13.16 | .07 | |
| 5186.82 | .06 | 30.30 | 6.50 | 13.33 | .07 | | 5366.64 | .03 | 4.70 | 4.50 | 12.83 | .13 | |
| 5188.36 | .07 | 25.10 | 2.60 | 13.92 | .07 | | 5367.08 | .02 | 3.50 | .80 | 13.18 | .14 | |
| 5189.09 | .12 | 14.50 | 5.80 | 13.21 | .22 | | 5367.37 | .01 | 2.40 | .50 | 13.73 | .21 | |
| 5195.10 | .02 | 7.00 | 2.80 | 12.83 | .07 | | 5367.62 | .01 | 2.40 | 1.10 | 13.86 | 1.06 | |
| 5205.90 | .05 | 23.10 | 3.90 | 13.24 | .06 | | 5368.86 | .04 | 34.40 | 3.20 | 14.30 | .05 | |
| 5208.99 | .03 | 9.50 | 3.70 | 12.84 | .07 | | 5370.54 | .04 | 34.60 | 2.90 | 14.17 | .04 | |
| 5223.86 | .04 | 20.10 | 3.80 | 13.34 | .06 | | 5372.90 | .06 | 55.30 | 5.60 | 13.50 | .03 | |
| 5224.72 | .04 | 15.70 | 3.30 | 13.26 | .06 | | 5374.37 | .03 | 29.40 | 3.20 | 13.54 | .03 | |
| 5225.60 | .05 | 5.00 | 7.40 | 12.68 | .18 | ZnII 2062 $z = 1.53$ | 5376.06 | .07 | 41.00 | 4.20 | 13.83 | .05 | |
| 5226.66 | .03 | 35.00 | 2.20 | 14.04 | .03 | | 5376.95 | .05 | 16.80 | 2.80 | 13.90 | .09 | |
| 5229.48 | .04 | 23.70 | 2.40 | 14.00 | .05 | | 5377.79 | .03 | 29.20 | 2.20 | 14.06 | .03 | |
| 5230.74 | .05 | 31.20 | 4.50 | 14.17 | .05 | | 5379.38 | .09 | 32.70 | 7.50 | 13.32 | .08 | |
| 5231.54 | .02 | 11.80 | 3.60 | 13.87 | .24 | | 5382.35 | .02 | 22.80 | 1.50 | 13.72 | .03 | |
| 5236.75 | .09 | 29.80 | 7.20 | 13.25 | .08 | | 5385.66 | .08 | 25.40 | 8.60 | 12.97 | .10 | |
| 5238.00 | .03 | 14.00 | 2.30 | 13.42 | .06 | | 5390.81 | .05 | 23.10 | 4.50 | 13.20 | .06 | |
| 5239.08 | .04 | 36.50 | 5.20 | 13.87 | .04 | | 5392.52 | .05 | 38.60 | 3.30 | 14.17 | .03 | Lya $z = 3.43$ |
| 5240.27 | .02 | 3.50 | 1.00 | 14.92 | .95 | | 5394.13 | .04 | 25.20 | 3.60 | 15.03 | .31 | Lya $z = 3.43$ |
| 5241.32 | .06 | 48.10 | 5.60 | 14.29 | .04 | | 5395.57 | .05 | 37.20 | 2.00 | 14.17 | .04 | Lya $z = 3.43$ |
| 5242.42 | .03 | 15.00 | 2.80 | 13.69 | .11 | | 5397.57 | .07 | 28.00 | 4.90 | 13.18 | .07 | |
| 5247.36 | .05 | 29.30 | 3.10 | 13.97 | .05 | | 5399.22 | .07 | 47.50 | 5.60 | 13.49 | .04 | |
| 5248.35 | .04 | 23.40 | 3.30 | 13.85 | .06 | | 5401.28 | .02 | 33.50 | 1.60 | 14.15 | .03 | |
| 5249.58 | .13 | 33.60 | 11.90 | 13.54 | .14 | | 5403.58 | .07 | 28.60 | 7.30 | 13.20 | .08 | |
| 5250.61 | .05 | 24.90 | 3.20 | 13.87 | .07 | | 5412.45 | .07 | 35.40 | 5.10 | 13.25 | .06 | |
| 5253.05 | .02 | 42.00 | 2.00 | 14.12 | .02 | | 5418.52 | .04 | 31.90 | 6.10 | 13.84 | .23 | |
| 5271.99 | .06 | 9.60 | 5.90 | 13.52 | .16 | CII 1334 $z = 2.95$ | 5419.67 | .05 | 18.40 | 5.50 | 12.39 | .08 | ZnI 2139 $z = 1.53$ |
| 5272.50 | .08 | 14.50 | 7.50 | 13.59 | .15 | CII 1334 $z = 2.95$ | 5422.70 | .04 | 22.60 | 2.60 | 13.30 | .04 | |
| 5274.64 | .02 | 16.20 | 1.80 | 13.58 | .04 | | 5432.30 | .03 | 11.00 | 9.70 | 13.04 | .06 | |
| 5276.23 | .03 | 31.70 | 2.10 | 13.82 | .03 | | 5434.97 | .02 | 25.40 | 1.70 | 13.80 | .03 | |
| 5279.48 | .03 | 27.30 | 3.10 | 13.69 | .04 | | 5440.29 | .08 | 37.20 | 7.00 | 13.34 | .07 | |
| 5280.49 | .03 | 17.30 | 2.20 | 13.49 | .05 | | 5441.93 | .03 | 35.90 | 3.20 | 13.81 | .03 | |
| 5281.84 | .03 | 9.00 | 3.00 | 12.93 | .08 | | 5443.23 | .05 | 27.00 | 4.40 | 13.52 | .05 | |
| 5283.75 | .03 | 19.80 | 2.40 | 13.36 | .04 | | 5445.05 | .04 | 36.30 | 3.00 | 14.23 | .04 | |
| 5285.68 | .03 | 22.60 | 2.40 | 13.52 | .04 | | 5446.58 | .04 | 31.10 | 3.90 | 13.93 | .04 | |
| 5288.20 | .05 | 42.70 | 4.80 | 13.98 | .03 | | 5447.75 | .05 | 21.10 | 4.10 | 13.35 | .07 | |
| 5289.82 | .03 | 44.40 | 2.70 | 14.14 | .03 | | 5449.92 | .01 | 33.10 | 1.40 | 14.38 | .04 | |
| 5291.75 | .03 | 47.50 | .90 | 15.21 | .04 | | 5451.79 | .04 | 31.20 | 3.50 | 13.30 | .04 | |
| 5295.92 | .06 | 19.80 | 6.00 | 13.03 | .09 | | 5453.04 | .03 | 26.60 | 2.90 | 13.30 | .03 | |
| 5298.22 | .02 | 31.00 | 1.60 | 14.11 | .04 | | 5454.79 | .04 | 22.20 | 3.20 | 13.05 | .05 | |
| 5313.62 | .05 | 32.50 | 4.30 | 13.39 | .05 | | 5456.10 | .03 | 16.20 | 3.30 | 13.18 | .05 | |
| 5316.82 | .02 | 22.40 | 1.90 | 13.47 | .03 | | 5461.24 | .02 | 21.60 | 2.10 | 13.54 | .03 | |
| 5317.60 | .02 | 16.90 | 2.00 | 13.62 | .03 | | 5463.34 | .02 | 27.80 | 2.00 | 13.74 | .03 | |
| 5318.67 | .02 | 29.00 | 1.60 | 13.85 | .02 | | 5470.72 | .02 | 34.90 | 2.00 | 14.06 | .02 | |
| 5322.45 | .07 | 33.80 | 3.80 | 13.96 | .07 | | 5472.17 | .05 | 12.90 | 4.20 | 13.03 | .11 | |
| 5323.86 | .05 | 56.30 | 3.20 | 14.36 | .03 | | 5473.07 | .02 | 26.90 | 1.70 | 13.94 | .02 | |
| 5327.06 | .02 | 28.50 | 1.40 | 14.20 | .04 | | 5475.88 | .02 | 30.60 | 1.30 | 13.80 | .02 | |

| $\lambda_{vac}$ | ± | $b$ | ± | $\log N$ | ± | ID |
|---|---|---|---|---|---|---|
| 5478.75 | .05 | 17.90 | 2.80 | 13.49 | .08 | |
| 5479.40 | .03 | 16.40 | 1.70 | 13.79 | .05 | |
| 5482.03 | .06 | 16.20 | 4.80 | 13.53 | .27 | |
| 5482.87 | .06 | 35.60 | 15.70 | 13.99 | .16 | |
| 5484.00 | .37 | 24.90 | 16.20 | 13.07 | .62 | |
| 5484.99 | .04 | 15.20 | 2.80 | 12.85 | .06 | |
| 5486.68 | .02 | 24.10 | 1.60 | 13.82 | .03 | |
| 5487.71 | .06 | 22.40 | 7.10 | 13.27 | .11 | |
| 5488.51 | .05 | 17.10 | 2.90 | 13.30 | .07 | |
| 5491.76 | .05 | 19.80 | 3.20 | 13.45 | .07 | |
| 5492.35 | .02 | 11.60 | 1.50 | 13.68 | .06 | |
| 5497.24 | .02 | 8.00 | 3.10 | 12.78 | .06 | |
| 5502.23 | .05 | 45.60 | .50 | 14.40 | .03 | |
| 5504.10 | .08 | 39.60 | 7.50 | 14.61 | .19 | |
| 5505.63 | .12 | 38.30 | 5.80 | 14.01 | .10 | |
| 5507.31 | .08 | 28.90 | 4.30 | 13.75 | .13 | |
| 5508.19 | .37 | 36.50 | 23.60 | 13.29 | .41 | |
| 5514.85 | .02 | 43.20 | 1.90 | 13.90 | .02 | |
| 5517.36 | .04 | 31.90 | 3.70 | 13.45 | .04 | |
| 5524.44 | .03 | 27.40 | 2.20 | 13.36 | .03 | |
| 5528.29 | .03 | 27.90 | 2.30 | 13.30 | .03 | |
| 5532.83 | .02 | 31.30 | 2.10 | 13.66 | .02 | |
| 5533.81 | .03 | 11.00 | 10.60 | 12.93 | .09 | |
| 5534.78 | .05 | 43.30 | 5.80 | 13.55 | .04 | |
| 5536.71 | .05 | 27.10 | .50 | 14.55 | .03 | |
| 5538.59 | .05 | 37.90 | .50 | 14.48 | .03 | |
| 5539.80 | .05 | 28.20 | 8.50 | 13.80 | .10 | |
| 5541.16 | .02 | 34.10 | 1.70 | 14.34 | .05 | |
| 5549.06 | .02 | 21.80 | 1.40 | 13.74 | .02 | |
| 5549.67 | .01 | 11.00 | .50 | 13.43 | .05 | |
| 5550.87 | .02 | 40.50 | 1.80 | 13.77 | .03 | |
| 5553.86 | .03 | 37.40 | 2.70 | 13.40 | .02 | |
| 5555.72 | .07 | 36.00 | 6.20 | 13.27 | .06 | |
| 5556.96 | .02 | 25.50 | 1.50 | 13.91 | .02 | |
| 5558.45 | .02 | 34.10 | 1.40 | 13.91 | .03 | |
| 5561.89 | .08 | 65.60 | 5.00 | 13.87 | .03 | |
| 5562.62 | .05 | 19.00 | 1.20 | 14.60 | .14 | |
| 5563.72 | .05 | 20.60 | 7.80 | 14.07 | .07 | |
| 5564.57 | .05 | 14.00 | 2.00 | 14.60 | .34 | |
| 5565.50 | .04 | 28.10 | 2.70 | 13.88 | .04 | |
| 5566.66 | .02 | 24.60 | 2.50 | 13.17 | .03 | |
| 5567.67 | .07 | 21.20 | 5.20 | 13.10 | .11 | |
| 5568.56 | .02 | 22.00 | 1.30 | 14.10 | .03 | |
| 5569.42 | .06 | 20.00 | 3.40 | 13.52 | .11 | |
| 5570.29 | .04 | 30.70 | 2.30 | 14.06 | .04 | |
| 5573.02 | .03 | 30.90 | 3.60 | 13.24 | .04 | |
| 5574.29 | .03 | 22.70 | 2.00 | 13.36 | .04 | |
| 5575.44 | .04 | 35.50 | 3.10 | 13.48 | .03 | |
| 5578.47 | .02 | 48.30 | 1.40 | 14.20 | .03 | |
| 5581.01 | .02 | 52.20 | 3.10 | 14.59 | .03 | |
| 5583.24 | .13 | 56.10 | 10.70 | 13.49 | .09 | |
| 5593.74 | .04 | 64.60 | 2.00 | 14.53 | .02 | Ly$\alpha$ $z = 3.60$ |
| 5596.15 | .03 | 26.80 | 1.50 | 16.50 | .29 | Ly$\alpha$ $z = 3.60$ |
| 5600.62 | .03 | 46.40 | 2.70 | 13.56 | .03 | |
| 5602.72 | .02 | 46.80 | 2.10 | 13.77 | .03 | |
| 5605.10 | .03 | 53.60 | 2.40 | 13.98 | .03 | |
| 5605.84 | .02 | 22.50 | 2.00 | 13.77 | .03 | |
| 5607.12 | .03 | 35.90 | 1.60 | 13.72 | .03 | |
| 5615.70 | .04 | 19.30 | 3.00 | 12.99 | .05 | |
| 5627.44 | .15 | 13.70 | 13.90 | 12.81 | .35 | |
| 5631.38 | .05 | 24.10 | 4.00 | 12.98 | .06 | |
| 5636.37 | .09 | 43.90 | 9.00 | 13.08 | .07 | |
| 5639.32 | .02 | 26.70 | 1.50 | 13.63 | .03 | |
| 5652.54 | .04 | 17.20 | 3.10 | 12.87 | .06 | |
| 5657.20 | .13 | 24.10 | 3.80 | 14.00 | .17 | |
| 5657.81 | .10 | 20.50 | 3.00 | 13.93 | .19 | |
| 5794.27 | .04 | 2.00 | .80 | 12.73 | .33 | SiII 1526 $z = 2.79$ |
| 5840.20 | .03 | 10.80 | 2.90 | 12.90 | .07 | SiIV 1393 $z = 3.19$ |
| 5841.14 | .00 | 18.10 | 4.80 | 12.89 | .08 | SiIV 1393 $z = 3.19$ |
| 5842.31 | .00 | 21.30 | 8.00 | 12.70 | .14 | SiIV 1393 $z = 3.19$ |
| 5845.58 | .07 | 14.10 | 6.00 | 12.59 | .12 | SiIV 1393 $z = 3.19$ |
| 5877.97 | .04 | 10.80 | 2.90 | 12.90 | .07 | SiIV 1402 $z = 3.19$ |
| 5878.92 | .00 | 18.10 | 4.80 | 12.89 | .08 | SiIV 1402 $z = 3.19$ |
| 5880.09 | .00 | 21.30 | 8.00 | 12.70 | .14 | SiIV 1402 $z = 3.19$ |
| 5913.81 | .03 | 3.10 | 7.50 | 13.01 | .68 | SiII 1526 $z = 2.87$ |
| 5939.06 | .00 | 8.80 | .70 | 13.72 | .05 | FeII 2344 $z = 1.53$ |
| 6015.69 | .00 | 8.80 | .70 | 13.72 | .05 | FeII 2374 $z = 1.53$ |
| 6030.92 | .05 | 10.50 | 3.80 | 12.98 | .09 | SiII 1526 $z = 2.95$ |
| 6031.61 | .05 | 1.90 | 1.90 | 12.69 | .24 | SiII 1526 $z = 2.95$ |
| 6036.73 | .00 | 8.80 | .70 | 13.72 | .05 | FeII 2382 $z = 1.53$ |
| 6116.09 | .00 | 11.90 | 3.40 | 13.32 | .08 | CIV 1548 $z = 2.95$ |
| 6116.53 | .00 | 8.90 | 3.80 | 13.16 | .10 | CIV 1548 $z = 2.95$ |
| 6126.26 | .00 | 11.90 | 3.40 | 13.32 | .08 | CIV 1550 $z = 2.95$ |
| 6126.70 | .00 | 8.90 | 3.80 | 13.16 | .10 | CIV 1550 $z = 2.95$ |
| 6325.52 | .05 | 29.20 | 8.50 | 13.34 | .10 | CIV 1548 $z = 3.08$ |
| 6336.04 | .05 | 29.20 | 8.50 | 13.34 | .10 | CIV 1550 $z = 3.08$ |
| 6340.77 | .05 | 12.00 | 2.50 | 12.69 | .05 | MgII 2796 $z = 1.26$ |
| 6341.50 | .00 | 11.90 | 1.80 | 12.80 | .04 | MgII 2796 $z = 1.26$ |
| 6357.04 | .04 | 12.00 | 2.50 | 12.69 | .05 | MgII 2803 $z = 1.26$ |
| 6357.78 | .00 | 11.90 | 1.80 | 12.80 | .04 | MgII 2803 $z = 1.26$ |
| 6469.16 | .09 | 9.30 | 6.40 | 11.68 | .17 | MgI 2852 $z = 1.26$ |
| 6469.94 | .18 | 12.80 | 20.90 | 11.46 | .39 | MgI 2852 $z = 1.26$ |
| 6487.15 | .03 | 14.00 | 1.70 | 13.77 | .04 | CIV 1548 $z = 3.19$ |
| 6488.21 | .00 | 20.10 | 2.10 | 14.06 | .03 | CIV 1548 $z = 3.19$ |
| 6489.67 | | | | | | CIV 1548 $z = 3.19$ |
| 6493.23 | .00 | 30.30 | 2.80 | 13.79 | .03 | CIV 1548 $z = 3.19$ |
| 6497.94 | .04 | 14.00 | 1.70 | 13.77 | .04 | CIV 1550 $z = 3.19$ |
| 6499.00 | .00 | 20.10 | 2.10 | 14.06 | .03 | CIV 1550 $z = 3.19$ |
| 6500.47 | | | | | | CIV 1550 $z = 3.19$ |
| 6504.03 | .00 | 30.30 | 2.80 | 13.79 | .03 | CIV 1550 $z = 3.19$ |
| 6553.27 | .00 | 8.80 | .70 | 13.72 | .05 | FeII 2586 $z = 1.53$ |
| 6587.53 | .00 | 8.80 | .70 | 13.72 | .05 | FeII 2600 $z = 1.53$ |
| 6588.82 | .03 | 15.70 | 4.20 | 13.31 | .08 | CIV 1548 $z = 3.25$ |
| 6599.78 | .04 | 15.70 | 4.20 | 13.31 | .08 | CIV 1550 $z = 3.25$ |
| 6600.46 | .05 | 9.90 | 4.20 | 12.07 | .11 | AlII 1670 $z = 2.95$ |
| 6869.00 | .00 | 15.30 | 4.50 | 13.41 | .09 | CIV 1548 $z = 3.43$ |
| 6880.43 | .00 | 15.30 | 4.50 | 13.41 | .09 | CIV 1550 $z = 3.43$ |

TABLE 3
Maximum likelihood analysis removing lines within 8 Mpc of the QSOs

| QSO sample | log $N$ range | number of lines | $\gamma$ | $\pm$ | $\beta$ | $\pm$ | P($\beta$) |
|---|---|---|---|---|---|---|---|
| 0055-269 | $\geq$13.3 | 178 | – | – | 1.80 | 0.03 | $7 \times 10^{-4}$ |
| 0055-269 | 13.3–14.5 | 170 | – | – | 1.38 | 0.08 | 0.44 |
| All | $\geq$13.3 | 507 | 2.17 | 0.41 | 1.74 | 0.04 | $2 \times 10^{-5}$ |
| All | 13.3–14.5 | 464 | 2.47 | 0.44 | 1.42 | 0.05 | 0.27 |
| All | 13.3–13.8 | 242 | 2.51 | 0.12 | 1.45 | 0.03 | 0.93 |
| All | $\geq$ 13.8 | 265 | 1.86 | 0.21 | 2.07 | 0.03 | 0.55 |